\documentclass[epj, referee]{svjour}
\usepackage{graphics, amssymb, amsfonts, amsmath}
\begin{document}
\title{Collision of one dimensional (1D) spin polarized Fermi gases in an optical lattice}
\author{Jussi Kajala\inst{1} \and Francesco Massel\inst{2} \and P\"{a}ivi T\"{o}rm\"a{} \inst{1,} \inst{3}}

%
%

\institute{Aalto University School of Science, P.O.Box 15100, FI-00076 Aalto, FINLAND \and Low Temperature Laboratory, Aalto University, P.O. Box 15100, FI-00076 Aalto, FINLAND \and Kavli Institute for Theoretical Physics, University of California, Santa Barbara, California 93106-4030, USA}
\date{\today}
%

\abstract{
In this work we analyze the dynamical behavior of the collision
between two clouds of fermionic atoms with opposite spin
polarization. By means of the time-evolving block decimation (TEBD)
numerical method, we simulate the collision of two one-dimensional
clouds in a lattice. There is a symmetry in the collision behaviour between
the attractive and repulsive interactions. We analyze the pair formation
dynamics in the collision region, providing a quantitative analysis of
the pair formation mechanism in terms of a simple two-site model. 
}
\maketitle
\section{Introduction}
\label{intro}

In recent years, ultracold gases have become an unparalleled
tool for simulating condensed matter systems \cite{Jaksch:2005p1471}
and to explore the properties of paradigmatic condensed matter models.
Ultracold gases allow an unprecedented tunability of the system parameters and
dimensionality. For instance, the experimental realization of the Fermionic Hubbard
Hamiltonian \cite{Schneider:2008p500,Joerdens:2008p276} represents an
important effort towards studying phenomena underlying high temperature
superconductivity and its connection with antiferromagnetism in
cuprates and iron pnictides. Ferromagnetic states in ultracold gases 
have been considered in the experiment \cite{Jo:2009p2226}.
Indeed, to understand the formation and properties of strongly correlated states
of Fermions it is essential to study also their dynamics. In condensed matter physics
much work has been devoted to the determination of the ground-state properties of the
Hubbard Hamiltonian (see e.g. \cite{HLEssler:2005p1368}), while its dynamical behavior has been
explored to a much lesser extent.
Nevetheless, in the recent past, the versatility of ultracold atomic
systems has led, from a numerical and theoretical point of view, to
approach the analysis of the dynamics in such systems
\cite{Kollath:2005p1574,Manmana:2007p100,Chang:2008p2230,Massel:2009p54,Kantian:2009p1477,Tezuka:2010p1573,Korolyuk:2010p1367,Joseph:2010p2229,Wall:2010p500},
leading to a revived interest in the unitary evolution of closed
quantum systems \cite{Rigol:2008p256}.  Recently, an interesting experimental investigation 
of spin dynamics in a system of colliding Fermi gas clouds was reported 
\cite{Sommer:2011p2205}, closely related to the topic of this article.


In this work, we simulate the collision of spin polarized gases using
the time-evolving block decimation (TEBD) algorithm
\cite{Vidal:2003p622}.  We are interested in the collisional
properties of two clouds with opposite spin polarization (denoted up
and down hereafter). Initially, the two polarized gases are trapped by
separate harmonic potentials. At $t=0$ we turn off the harmonic
traps, allowing the clouds to expand and collide with each other, in
complete analogy to what has been done in the experiment reported in
\cite{Sommer:2011p2205}, except for the presence of the lattice.

Both attractive and repulsive interactions between the species are
considered. Counterintuitively, the physics of the collision is independent
of the sign of the interaction. From an intuitive point of view, one
might expect that, in presence of attractive interaction, the
particles would merge in a gas of pairs, while bouncing off for
repulsive interaction. However, the actual quantum unitary dynamics 
is different.  Of particular interest is the pair creation during
the collision for both signs of the interaction strength. In the
present article, it is shown that the mechanism for pair creation in
the collision is explained by a two-site analysis previously introduced
by us \cite{Kajala:2011p500}. It is also shown that the qualitative dynamics of the
collision fundamentally arises from this Hubbard Dimer model.
Finally, we discuss the connection between our results and the
results in the low-temperature strongly
interacting regime of the recent experiment \cite{Sommer:2011p2205}.

In section \ref{system} we describe the system in detail. 
In section \ref{TEBD} we elaborate on the TEBD numerics and in section \ref{HDimer}
explain the Hubbard Dimer two-site model. Finally, in section \ref{results} we
compare the results of the numerics and the model.

\section{The system}
\label{system}

We consider one-dimensional spin up and spin down gases confined in separate potentials in space. In addition to the harmonic trapping
potentials, there is a lattice potential (see Fig. \ref{fig:schematic}). The system is described by the Hubbard Hamiltonian with a harmonic potential:

 \begin{figure}
      \resizebox{0.99\columnwidth}{!}{
	\includegraphics{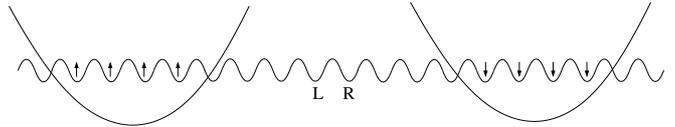}
      }
      \caption{The system. Spin up and down gases in a lattice are confined in two separate harmonic potentials. At time zero the harmonic potentials are removed
           	and the gases expand and collide. Here, L and R mark the two central sites where the expanding gases meet and which we use for the two-site model.} 
      \label{fig:schematic}   
 \end{figure}

 \begin{equation}
   \label{eq:HH}
   \begin{array}{ll}
     H_H=U \sum_i \hat{n}_{i ,\uparrow}\hat{n}_{i,\downarrow} 
     + V_{\uparrow} \sum_i (C_{\uparrow} - i)^2 \hat{n}_{i,\uparrow} \\
     + V_{\downarrow} \sum_i (C_\downarrow - i)^2 \hat{n}_{i,\downarrow}
     - J \sum_{i\,\sigma=\uparrow,\downarrow} c^\dagger_{i\,\sigma}c_{i+1\,\sigma} + h.c. ,
   \end{array}
\end{equation}
where the $\hat{c}^{\dagger}_{i, \downarrow}$ operator creates a spin down particle at lattice 
site i, $\hat{c}_{i, \uparrow}$ annihilates an up  particle at lattice site i, 
$\hat{n}_{i,\uparrow} = \hat{c}^{\dagger}_{i, \uparrow}  \hat{c}_{i, \uparrow}$,
$J$ is the hopping matrix element, $U$ is the interaction strength between the spin up 
and down particles, and $V$ and $C$ are the spin-dependent harmonic trapping strength and
position of the trap center, respectively. 

The initial state of the simulations is the ground state of the Hubbard Hamiltonian \eqref{eq:HH} with $V_{\uparrow} = V_{\downarrow} = 0.02$, $C_{\uparrow} = 50.5$, and 
$C_{\downarrow} = 100.5$.
Then, this state is evolved with the otherwise same Hamiltonian, except $V(i)_{\uparrow} = V(i)_{\downarrow} = 0.0$.

\section{TEBD numerics}
\label{TEBD}

In this work, we use the time-evolving block decimation (TEBD) algorithm \cite{Vidal:2003p622} to
model the collision in one dimension and in a lattice. The only approximations involved in the TEBD numerics are in the 
Suzuki-Trotter expansion and in the Schmidt truncation of the Hilbert space \cite{Vidal:2003p622}. TEBD is an essentially exact numerical method in the sense that it does not 
use mean-field approximations, and the errors due to the approximations above are controllable.
For the details of the TEBD numerics, see \cite{Vidal:2003p622}. 

The initial state for the time-evolution is calculated using a ground state algorithm for TEBD \cite{Vidal:2003p622}. We have used two different algorithms, one
for the time evolution and another for the ground state, and both of the algorithms employ TEBD. The ground state calculating algorithm is called the ''imaginary time
evolution algorithm'' and the time evolution one is called the ''real time evolution algortithm''. The former solves the ground state of the Hubbard Hamiltonian
and the latter determines the time evolution when the wavefunction is acted on by $e^{-i \hat{H} t}$, i.e. by solving the time-dependent Schr\"{o}dinger's equation.
The Schmidt number for TEBD (describing the numerical truncation) in the simulations is $\Gamma = 150$.

For the simulation parameters we choose to have 20 up and 20 down particles, i.e. $N_{\downarrow} = 20$, $N_{\uparrow} = 20$.The interaction is varied so that 
we run the simulation for interactions $\frac{U}{J} = 0, \pm 1, \pm 3, \pm 5, \pm 7, \pm 10, \pm 15$ and $\pm 20$, where the interaction is expressed
in the units of the hopping $J$. Above and from now on, variables are expressed in the units of hopping, and $J = 1$ has been chosen in the numerics.
Note that in our convention negative values of $U$ represent attractive interaction (see Equation \eqref{eq:HH}).
The initial trapping strengths are chosen to be $\frac{V}{J} = 0.02$. We consider the temperature $T = 0$.
At time zero, we release the traps (i.e. change $V$ from $0.02 J$ to $0.00 J$) and let the two clouds expand, keeping the interactions on during the expansion. 

We run the time evolutions up to the time $t = 25 \frac{1}{J}$, as then the outer edges of the clouds have hit the edges of our finite system 
(the lattice size $L = 150$), and we are not interested in the unphysical edge collision dynamics. However, the maximum speed for propagation of 
the distrurbance caused by the collision with the edge is $2J$ 
(due to the lattice dispersion). As we are interested only in what happens in the collision center,
we can run the simulations a little longer than the time when collision with the edges occurs.
 
As a result of the TEBD simulations, we obtain the density profiles of up and down particles, $n_{i, \uparrow} (t)$, $n_{i, \downarrow} (t)$. Importantly,
also the density of doublons, $n_{i, \uparrow \downarrow} (t)$ is obtained:

\begin{equation}
  \begin{array}{ll}
\label{eq:pairdensity}

n_{i\, \uparrow \downarrow} (t) = < \Phi (t) | c^\dagger_{i\, \uparrow} c_{i\,\uparrow} c^\dagger_{i\, \downarrow}  c_{i\, \downarrow} |\Phi (t)  >,
  \end{array}
\end{equation}
where $<>$ denotes the quantum mechanical expectation value, and $\Phi (t)$ is the wavefunction. Elaborating on the definition of doublons,
the doublons are excitations of the form  $c_{i\,\uparrow}^\dagger c_{i\, \downarrow}^\dagger|\emptyset \rangle$ and the single (unpaired) particles are defined as
$c_{i\,\sigma}^\dagger|\emptyset\rangle$ ($\sigma=\uparrow,\,\downarrow$), where $|\emptyset \rangle$ is the state representing an empty lattice site. 
The local number of doublons is  given by $n_{i\, \uparrow \downarrow} (t)$, 
while the number of unpaired (up) particles is given by  $n_{i\, \uparrow}^{un} (t) =  n_{i\, \uparrow} (t) - n_{i\,\uparrow \downarrow} (t)$. 
Now, before moving on to the results of the simulations let us discuss the theoretical model that we employ in order to explain the results. 

\section{The Hubbard Dimer two-fluid model}
\label{HDimer}

Previously, we have developed a Hubbard Dimer two-fluid model to explain the dynamics of expansion in a 1D Fermi gas in a lattice \cite{Kajala:2011p500} . 
There we considered the expansion of an interacting two-component gas which was initially set into a band insulator state, related to the experiment of \cite{Schneider:2010p1468}. As will be shown in this article, the same model explains dynamics of the collision of 1D polarized cases.
Below, we will go through the derivation of the Hubbard Dimer model in the 
case relevant for this problem (it is basically the same analysis as done in the online supporting material of \cite{Kajala:2011p500} but is included here, 
and done in greater detail, for clarity).

We assume that the important dynamics occurs in the collision center, i.e. the two lattice sites in the center where the expanding polarized gases meet, 
see Fig \ref{fig:schematic}. The spin basis for a single lattice site is $|\emptyset>$ (empty lattice site), 
$|\uparrow>$, $|\downarrow>$, and $|\uparrow \downarrow>$. 
Let us now assume that in the two-site collision center, just after the first particles have entered the system from the rest of the chain, we have the
state $|\Phi(t=0)> = |\uparrow , \downarrow>$. We want to determine how this state evolves into a doublon state, 
$|\uparrow \downarrow, \emptyset>$ or $|\emptyset, \uparrow \downarrow>$ as a function of time. In order to do that, we solve the two-site system with 
the Hubbard Hamiltonian exactly by diagonalizing it. The two site system and its solution is in general called the Hubbard Dimer (see e.g. \cite{Trotzky:2008p2201}). 

We have a 2-particle basis, and the Hubbard Hamiltonian conserves the number of particles. 
Due to anticommutation relations, a given order for the application of fermionic operators must be chosen:
    
    \begin{eqnarray}
      \label{eq:basis}
      | \uparrow, \downarrow>=c^\dagger_{1\,\uparrow} c^\dagger_{2\, \downarrow} |0>, \quad 
      | \downarrow, \uparrow>=c^\dagger_{1\, \downarrow} c^\dagger_{2\,\uparrow } |0>, \\
      | \uparrow \downarrow,0>=c^\dagger_{1\,\uparrow} c^\dagger_{1\, \downarrow} |0>, \quad
      | 0, \uparrow \downarrow>=c^\dagger_{2\,\uparrow} c^\dagger_{2\, \downarrow} |0> .
    \end{eqnarray}
The Hamiltonian is

    \begin{eqnarray}
      \label{eq:HubbHam}
        H=H_J+H_{int} \nonumber \\
       H_J=-J \sum_\sigma c^\dagger_{1,\sigma} c_{2\,\sigma} + h.c. \nonumber  \\
       H_{int}= U\sum_{i=1,2} n_{i\,\uparrow}n_{i\,\downarrow}.
    \end{eqnarray}
Then

    \begin{eqnarray}
      \label{eq:Hopp1}
      H_J | \uparrow, \downarrow>&=& H_J\, c^\dagger_{1\,\uparrow} c^\dagger_{2\, \downarrow} |0> \nonumber \\
       &=& -J\left(
                c^\dagger_{1\,\downarrow} c_{2\, \downarrow}
                c^\dagger_{1\,\uparrow} c^\dagger_{2\, \downarrow} |0> +
                c^\dagger_{2\,\uparrow} c_{1\, \uparrow}
                c^\dagger_{1\,\uparrow} c^\dagger_{2\, \downarrow} |0>  
          \right) \nonumber \\
     &=& J\left( | \uparrow \downarrow,0> - | 0, \uparrow \downarrow> \right),
    \end{eqnarray}

    \begin{eqnarray}
      \label{eq:Hopp2}
      H_J | \downarrow, \uparrow>&=& H_J\, c^\dagger_{1\,\downarrow} c^\dagger_{2\, \uparrow} |0> \nonumber \\
       &=& -J\left(
                c^\dagger_{1\,\uparrow} c_{2\, \uparrow}
                c^\dagger_{1\,\downarrow} c^\dagger_{2\, \uparrow} |0> +
                c^\dagger_{2\,\downarrow} c_{1\, \downarrow}
                c^\dagger_{1\,\downarrow} c^\dagger_{2\, \uparrow} |0>  
          \right) \nonumber \\
     &=& J\left(| 0, \uparrow \downarrow> -  | \uparrow \downarrow,0> \right)
    \end{eqnarray}
    and, obviously,
 
    \begin{eqnarray}
      \label{eq:Hopp3}
      H_J | \uparrow \downarrow, 0>&=& H_J\, c^\dagger_{2\,\uparrow} c^\dagger_{1\, \uparrow} |0> \nonumber \\
       &=& -J\left(
                c^\dagger_{2\,\uparrow} c_{1\, \uparrow}
                c^\dagger_{1\,\uparrow} c^\dagger_{1\, \downarrow} |0> +
                c^\dagger_{2\,\downarrow} c_{1\, \downarrow}
                c^\dagger_{1\,\uparrow} c^\dagger_{1\, \downarrow} |0>  
          \right) \nonumber \\
     &=& J\left( | \uparrow, \downarrow> - | \downarrow, \uparrow> \right),
    \end{eqnarray}

     \begin{eqnarray}
      \label{eq:Hopp4}
      H_J |0, \uparrow \downarrow>&=& H_J\, c^\dagger_{2\,\uparrow} c^\dagger_{2\, \uparrow} |0> \nonumber \\
       &=& -J\left(
                c^\dagger_{1\,\uparrow} c_{2\, \uparrow}
                c^\dagger_{2\,\uparrow} c^\dagger_{2\, \downarrow} |0> +
                c^\dagger_{1\,\downarrow} c_{2\, \downarrow}
                c^\dagger_{2\,\uparrow} c^\dagger_{2\, \downarrow} |0>  
          \right) \nonumber \\
     &=& J\left( | \uparrow, \downarrow> - | \downarrow, \uparrow> \right).
    \end{eqnarray}
    Hence in the 4-dimensional Hilbert space of the $N=2$ particles
    Hubbard Dimer, with the choice of the basis given by Eq.
    \eqref{eq:basis} (representation of $H_{int}$ is trivial), $H$ has the following representation 
    \begin{equation}
      \label{eq:H_I_matr1}
      \mathcal{\mathbf{H}}=\left[ 
         \begin{array}{cccc}
           0 & 0  &  J & -J  \\
           0 & 0  & -J &  J  \\
           J & -J &  U &  0  \\
           -J &  J &  0 &  U    
         \end{array}
      \right].
    \end{equation}
    The Hamiltonian can be rewritten in a basis where it
    assumes a block-diagonal form 

     \begin{equation}
      \label{eq:H_I_matr2}
      \mathcal{\mathbf{H_{bl}}}=\left[ 
         \begin{array}{cc|cc}
           0 & 0  &  0   &  0   \\
           0 & U  &  0   &  0   \\
          \hline
           0 & 0  &  0   &  -2J \\
           0 & 0  &  2J  &   U   
         \end{array}
      \right]
      ,
    \end{equation}   
    analogously, the hopping ``perturbation'' part of the Hamiltonian assumes the form
    \begin{equation}
      \label{eq:H_I_matr3}
      \mathcal{\mathbf{H_{J,bl}}}=\left[ 
         \begin{array}{cc|cc}
           0 & 0  &  0   &  0   \\
           0 & 0  &  0   &  0   \\
          \hline
           0 & 0  &  0   &  -2J \\
           0 & 0  &  2J  &   0   
         \end{array}
      \right].
    \end{equation}
    This representation corresponds to the following basis vectors
    \begin{eqnarray}
      \label{eq:symmEig}
      &&|T>  =\frac{1}{\sqrt{2}}\left( | \uparrow, \downarrow> + | \downarrow, \uparrow> \right) 
            \nonumber \\
      &&|D_->=\frac{1}{\sqrt{2}}\left( | \uparrow \downarrow,0> - | 0, \uparrow \downarrow> \right)
            \nonumber \\
      &&|S>  =\frac{1}{\sqrt{2}}\left( | \uparrow, \downarrow> - | \downarrow, \uparrow> \right) 
            \nonumber \\
      &&|D_+>= \frac{1}{\sqrt{2}}\left( | \uparrow \downarrow,0> + | 0, \uparrow \downarrow> \right).
    \end{eqnarray}
    If the lower block of $\mathcal{\mathbf{H}}$ is diagonalized, one
    obtains the following expression for the eigenvalues
    \begin{equation}
      \label{eq:eig}
      \lambda_{\pm}=U/2\left[1 \pm \sqrt{1+\frac{16 J^2}{U^2}} \right].
    \end{equation}
    Defining 
    \begin{equation}
      \label{eq:alpha}
      \alpha_{\pm}=-\frac{U}{4J}\left[1 \pm \sqrt{1+\frac{16 J^2}{U^2}}\right]
    \end{equation}
    the eigenvectors can be written as
    \begin{equation}
      \label{eq:eigv}
      |v_{\pm}>=\frac{1}{\sqrt{1+\alpha_{\pm}^2}}\left(|S>+\alpha_{\pm}|D_+>\right).
    \end{equation}
     Hence the full spectrum of the dimer is given by
    \begin{equation}
      \label{eq:spectr}
      \begin{array}{ll}
      \lambda_-=U/2\left[1 - \sqrt{1+\frac{16 J^2}{U^2}} \right] (<0) \qquad &\Leftrightarrow
      \qquad |v_-> \\
      \lambda_0=0    \qquad &\Leftrightarrow \qquad  |T>  \\
      \lambda_U=U    \qquad &\Leftrightarrow \qquad  |D_-> \\
      \lambda_+=U/2\left[1 + \sqrt{1+\frac{16 J^2}{U^2}} \right] (>U) \qquad &\Leftrightarrow
      \qquad |v_+>.
     \end{array}
     \end{equation}
    Now, in the case of the problem in hand we have initially the state
    
    \begin{equation}
      \label{eq:initial}
      |\phi (t=0)>  = | \uparrow, \downarrow>= \frac{1}{\sqrt{2}}(|S> + |T>)
    \end{equation}
    which we need to express in eigenstates of the Hamiltonian. Expressing $|S>$ as a superposition of $|v_+>$ and $|v_->$ gives:

    \begin{equation}
      \label{eq:initial_eigenstates}
      |\uparrow, \downarrow>= \frac{1}{\sqrt{2}}(\theta_- |v_-> - \theta_+ |v_+> + |T>),
    \end{equation}
    where we have denoted 
       
    \begin{equation}
      \label{eq:gamma}
      \theta_{\pm} = \frac{(\sqrt{1+\alpha_{\pm}^2}) \alpha_\mp}{(\alpha_{+} - \alpha_{-})} .
    \end{equation}
    Now let us determine the number of doublons in the left site, 
    $ n^{L}_{\uparrow \downarrow} (t) = <\phi (t) | \hat{n}^{L}_{\uparrow \downarrow} | \phi (t)> $ given by

    \begin{equation}
      \label{eq:time-evo}
      <\uparrow, \downarrow|  e^{\frac{i\hat{H}t}{\hbar}} \,  \hat{n}^{L}_{\uparrow \downarrow} \, e^{-\frac{i\hat{H}t}{\hbar}}  |\uparrow, \downarrow > .
    \end{equation}
    Calculating this gives
    
    \begin{equation}
      \label{eq:HDIMER}
      \begin{array}{ll}
	n^L_{\uparrow \downarrow} (t) =  \frac{8}{16 + \frac{U^2}{J^2}} [1 - cos(\sqrt{U^2 + 16J^2}t)],
      \end{array}
    \end{equation}
    which determines the time dependence of the doublons in the problem, completing the analysis. 

    Now, we need to return to our initial assumptions. Equation \eqref{eq:HDIMER} holds for the initial state $|\Phi(t=0)>^C = |\uparrow , \downarrow>$.
    If more unpaired particles did not enter the system, Equation \eqref{eq:HDIMER} would predict that we simply see oscillations
    in the doublon density with the frequency $\sqrt{U^2 + 16J^2}$ and amplitude $ \frac{8}{16 + \frac{U^2}{J^2}}$. 
    However, when the two gases collide, during the collision more unpaired particles enter the system, going to the state
    $|\uparrow , \downarrow>$. We make the hypothesis that the number of doublons in the central sites $n^C_{\uparrow \downarrow} (t)$ can be determined
    by the following short time approximation:
    \begin{equation}
      \label{eq:incoherent}
      \begin{array}{ll}
	n^C_{\uparrow \downarrow} (t) =  \int_{\tau = 0}^{\tau = t}  \int_{t^{'} = t}^{t^{'} = t}
	2 * \frac{8}{16 + \frac{U^2}{J^2}} \\
	* [1 - cos(\sqrt{U^2 + 16J^2}(\tau-t^{'}))] * sin^{2}(J*t^{'}) \\
	* n^{N}_{un}(t^{'}) d\tau dt^{'},
      \end{array}
    \end{equation}
    where C denotes the two-site collision center. Equation \eqref{eq:incoherent} contains the time evolution of pairs given by the Dimer problem, Equation
    \eqref{eq:HDIMER}. In addition, it takes into account the number of unpaired particles at the sites surrounding the two central sites, $n^{N}_{un}(t^{'})$,
    which can change during the collision. The particles from these neighbouring sites tunnel into the central two sites as described by the term $sin^{2}(J*t^{'})$.
    Particles tunnelled into the central two sites at time $t^{'}$ start the Dimer dynamics at that time, thus the shift $t-t^{'}$ in the cosine term. The factor
    of two comes from spatial symmetry.
    
    Finally, let us discuss the limitations of the above analysis.
    Equation \eqref{eq:incoherent} we assume to hold for short times, since for longer times one needs to take into account 
    unpaired particles as well as pairs tunnelling out of the two-site dimer.
    Incorporating these two into the analysis is somewhat meticulous, 
    but as we are, for now, more interested in whether the dynamics is fundamentally explained by the Hubbard Dimer model than whether the model dynamics 
    can be analytically solved at long times, we restrict ourselves to the short time limit. 
    The short time limit means the times when the change in $n^{N}_{un}(t)$ due to tunnelling into the central sites and
    due to pair creation is small, and the pair tunnelling is neglibile.
    The change in $n^{N}_{un}(t)$ is small when the $sin^{2}(J*t) = \frac{1 - 2 cos(2Jt)}{2}$ is close to zero. That holds
    when $t << \frac{\pi}{2} \frac{1}{J}$. This is shorter timescale than the pair tunnelling timescale for all $U$ and therefore is the limiting timescale in
    our short time analysis.

    This completes our analysis of the Hubbard Dimer. Let us next use the obtained analytical results to explain the numerics. \\

\section{Results and Discussion}
\label{results}

\begin{figure}
      \resizebox{0.99\columnwidth}{!}{
	\includegraphics{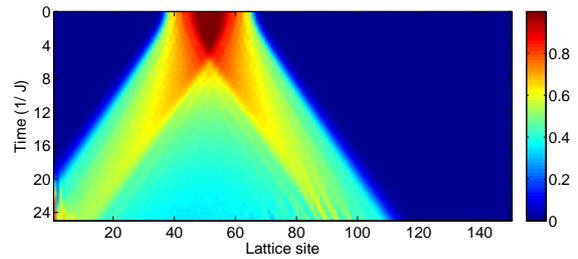}
      }
      \caption{The square root of the density profile of up particles, $\sqrt{n_{\uparrow, i} (t)}$, the interaction $U= 0.0$. $\sqrt{n_{\downarrow, i} (t)}$ is not shown
      as it is symmetric to $\sqrt{n_{\uparrow, i} (t)}$ with respect to the central lattice sites 76-77. (See online for colour).}
      \label{fig:U0_imagesc}   
    \end{figure}

\begin{figure}
      \resizebox{0.99\columnwidth}{!}{
	\includegraphics{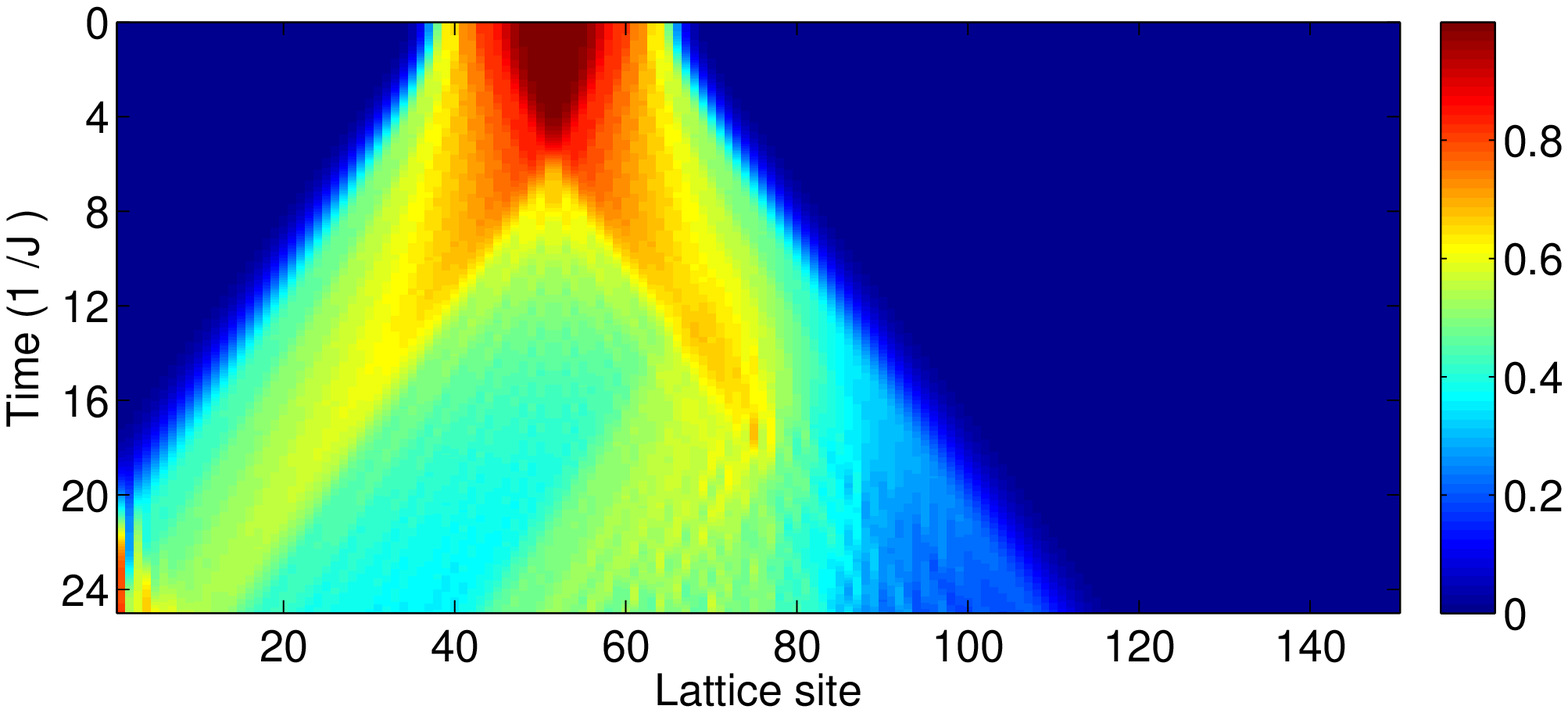}
      }
      \caption{The same as figure \ref{fig:U0_imagesc}, but $U = -1.0$.}
      \label{fig:U1_imagesc}   
    \end{figure}

\begin{figure}
      \resizebox{0.99\columnwidth}{!}{
	\includegraphics{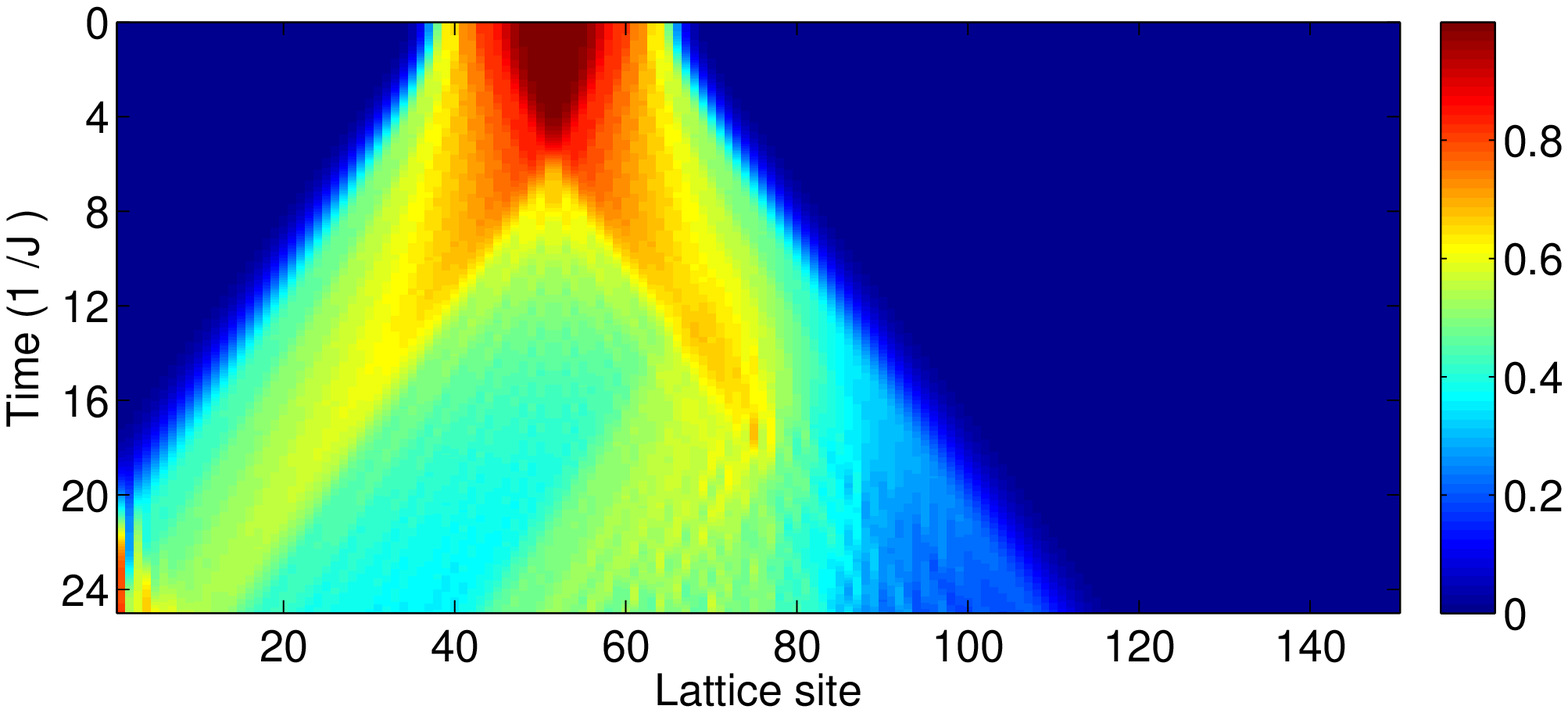}
      }
      \caption{The same as figure \ref{fig:U0_imagesc}, but $U = +1.0$.}
      \label{fig:U1p_imagesc}   
    \end{figure}

\begin{figure}
      \resizebox{0.99\columnwidth}{!}{
	\includegraphics{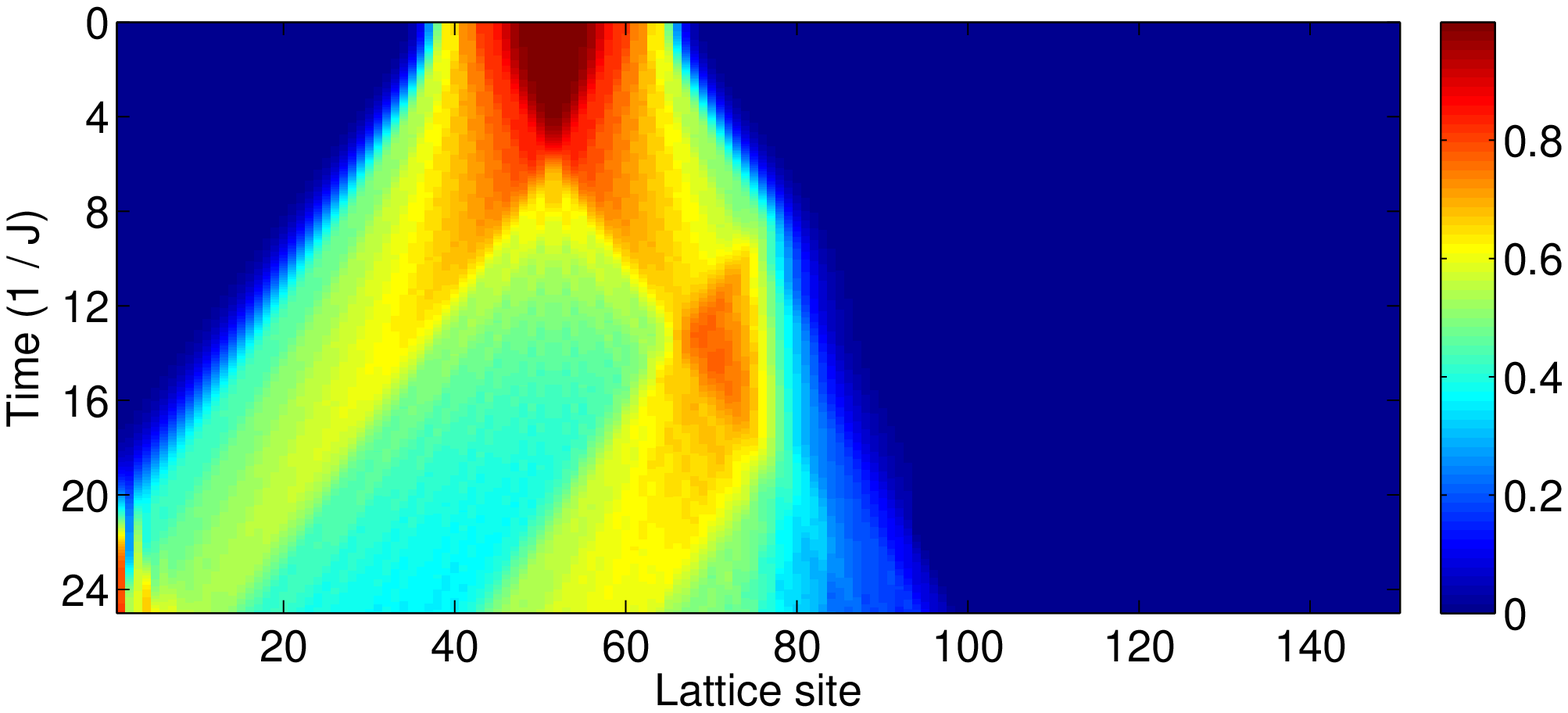}
      }
      \caption{The same as figure \ref{fig:U0_imagesc}, but $U = -5.0$.}
      \label{fig:U5_imagesc}   
    \end{figure}

\begin{figure}
      \resizebox{0.99\columnwidth}{!}{
	\includegraphics{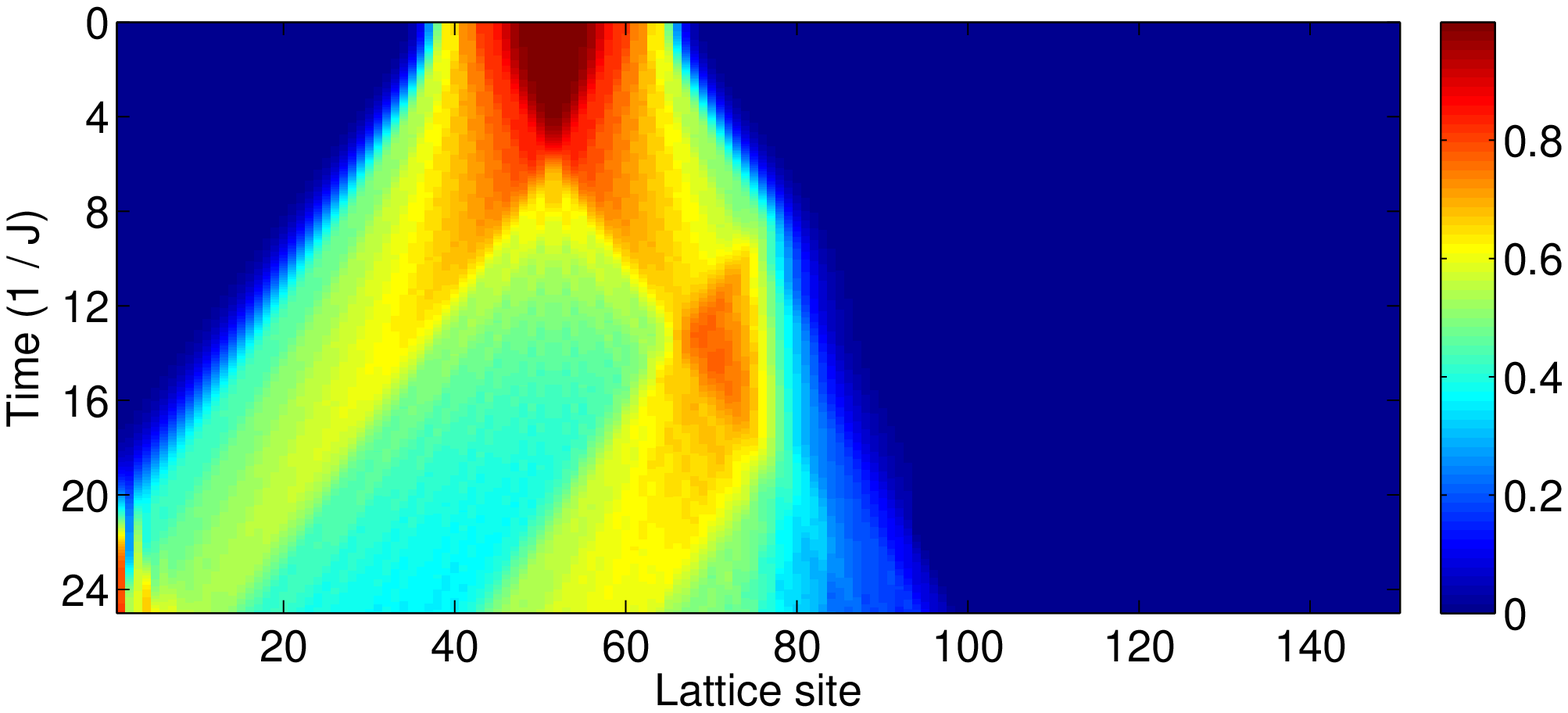}
      }
      \caption{The same as figure \ref{fig:U0_imagesc}, but $U = +5.0$.}
      \label{fig:U5p_imagesc}   
    \end{figure}

\begin{figure}
      \resizebox{0.99\columnwidth}{!}{
	\includegraphics{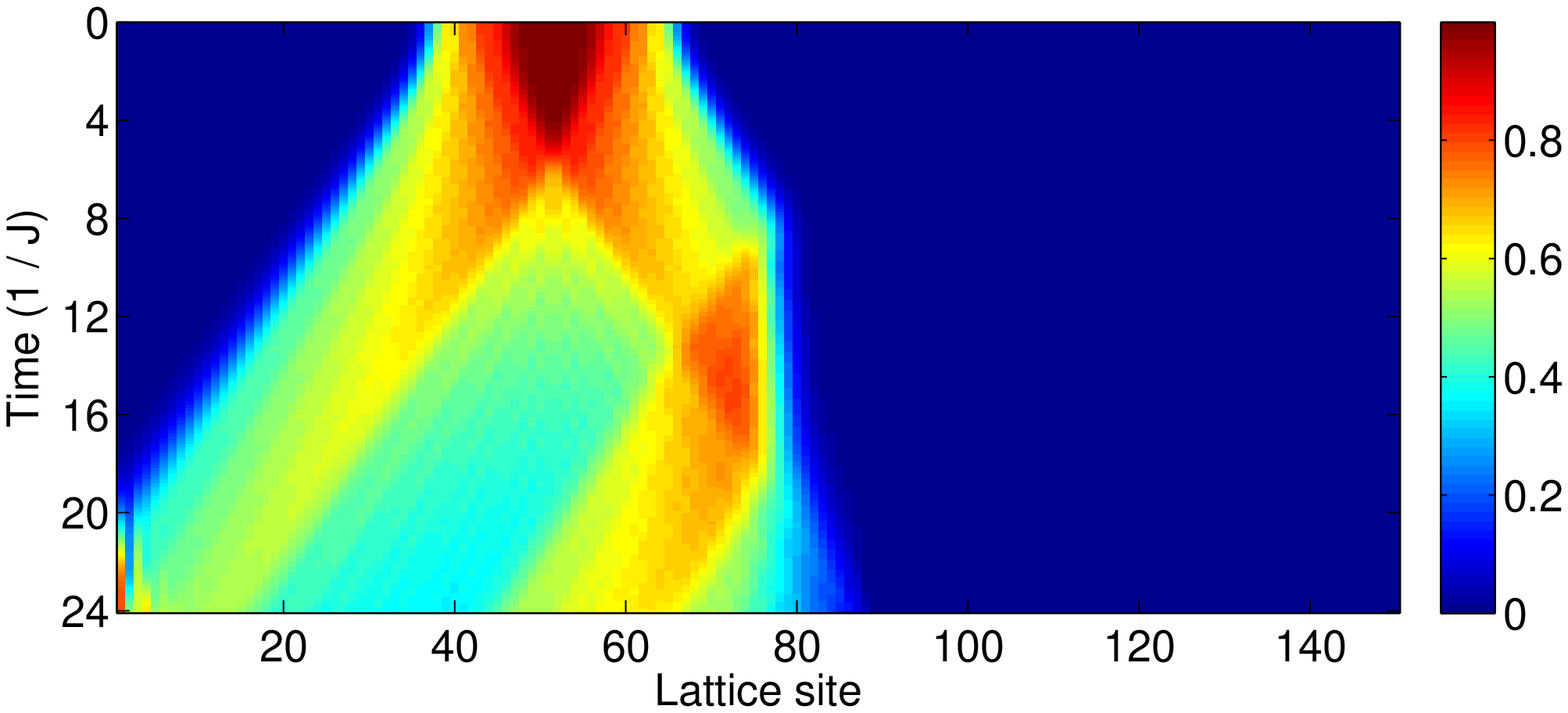}
      }
      \caption{The same as figure \ref{fig:U0_imagesc}, but $U = -15.0$.}
      \label{fig:U15_imagesc}   
    \end{figure}

    The square roots of the up density profiles $\sqrt{n_{\downarrow, i} (t)}$ during the collision of the oppositely polarized gases
    are shown in Figures \ref{fig:U0_imagesc} - \ref{fig:U15_imagesc} for different interactions. We are plotting the square roots of the density
    distributions since they highlight low density features which are important for the analysis in the case of pairs. Note that due to symmetry the density profiles
    of down particles $\sqrt{n_{\uparrow, i} (t)}$ are mirror images of $\sqrt{n_{\uparrow, i} (t)}$ with respect to the collision center.
    
    \begin{figure}
      \resizebox{0.99\columnwidth}{!}{
	\includegraphics{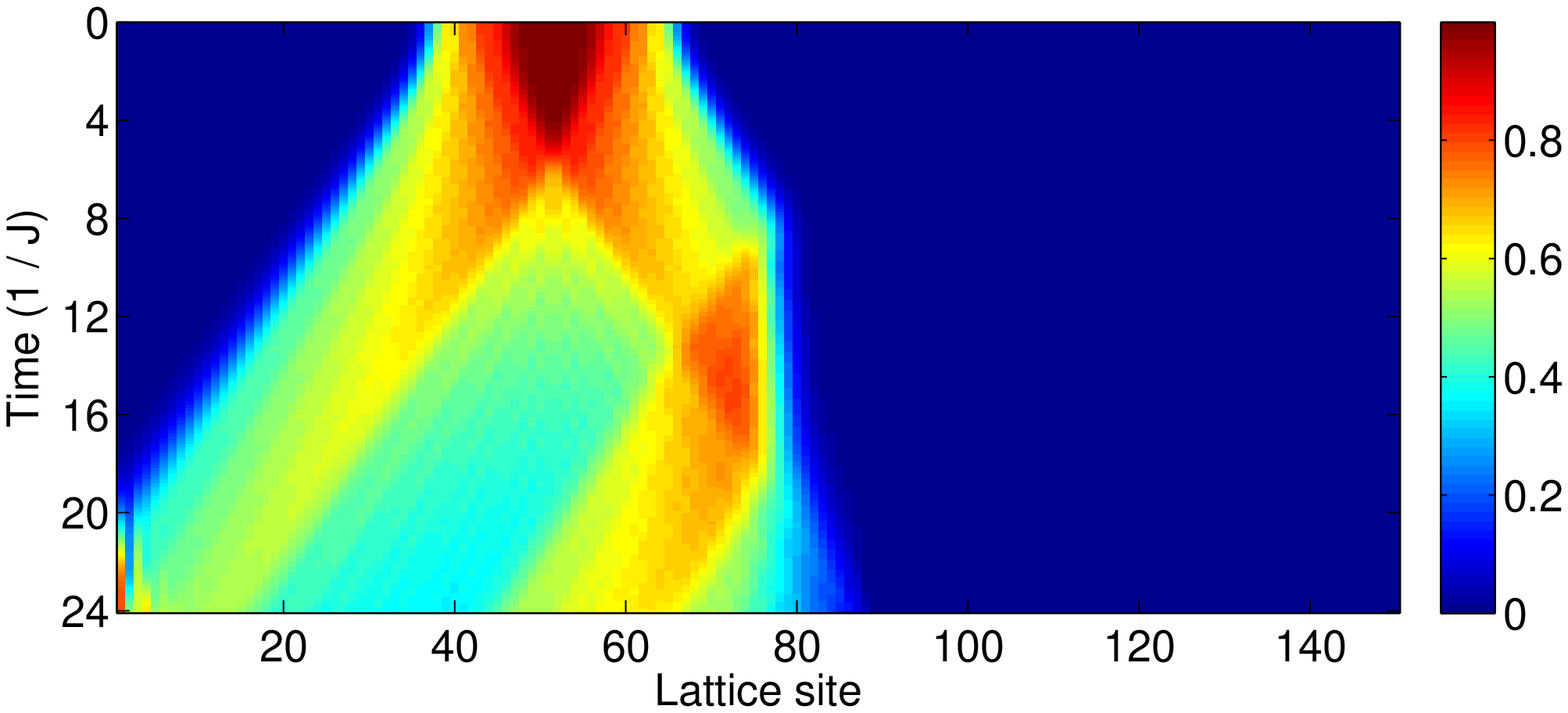}
      }
      \caption{The same as figure \ref{fig:U0_imagesc}, but $U = +15.0$.}
      \label{fig:U15p_imagesc}   
    \end{figure}
    Looking at Figures \ref{fig:U0_imagesc} - \ref{fig:U15p_imagesc} one interestingly observes at every interaction $|U|$ that there is a $U \leftrightarrow -U$ symmetry 
    in the collision. The symmetry holds for all observables we determined. In the case of both the attractive and repulsive
    interactions the clouds bounce back from each other. One way of explaining, for large $|U|$, this somewhat surprising behaviour is noting that the lattice dispersion 
    limits possible kinetic energy in the single band Hubbard model, the maximum energy being $4J$. 
    When $|U|$ is large, the large energy mismatch between a paired state and a non-paired one suppresses the probability of creating a pair from the colliding initially
    unpaired clouds, as reflected in the Lorentzian form of the amplitude $\frac{1}{2}\frac{1}{1+ \frac{U^2}{16 J^2}}$ in Equation \eqref{eq:HDIMER}.
    Therefore the polarized clouds are reflected from each other. 
    Below, we shall compare the amount of doublons created in numerics to the predictions
    of the Hubbbard Dimer. Noting the $U \leftrightarrow -U$ symmetry we will henceforth in the discussion denote interactions with absolute values.

\begin{figure}
      \resizebox{0.99\columnwidth}{!}{
	\includegraphics{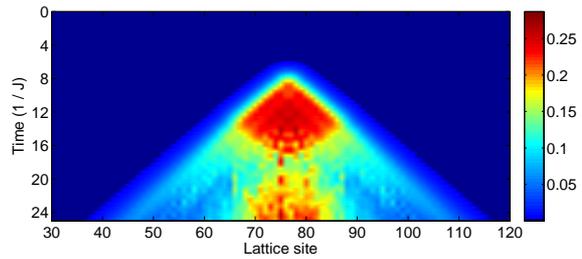}
      }
      \caption{The square root of the density profile of doublons, $\sqrt{n_{\uparrow, i} (t)}$, the interaction is $|U| = 1.0$. (See online for colour).}

      \label{fig:pU1_imagesc}   
    \end{figure}

\begin{figure}
      \resizebox{0.99\columnwidth}{!}{
	\includegraphics{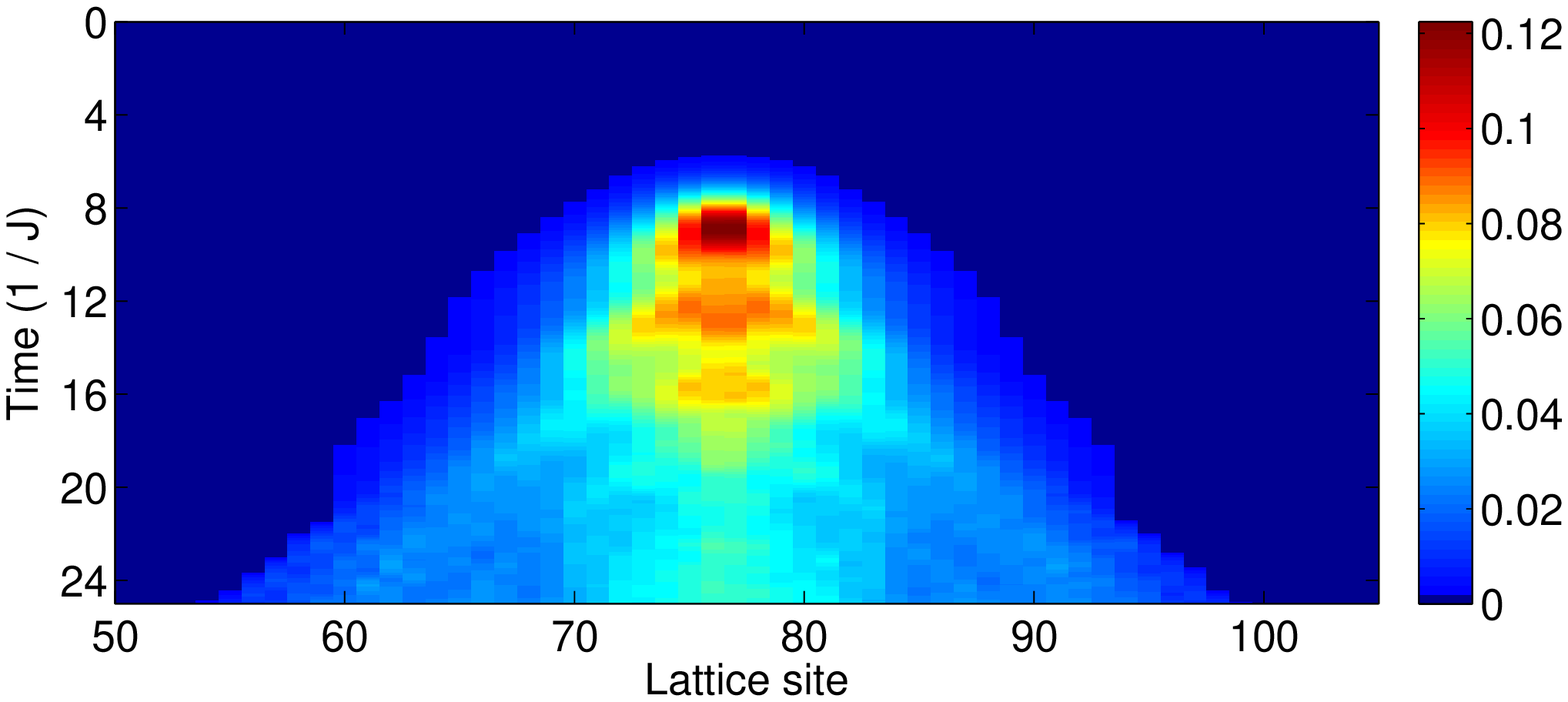}
      }
      \caption{The same as figure \ref{fig:pU1_imagesc}, but $|U| = 5.0$.}
      \label{fig:pU5_imagesc}   
    \end{figure}

\begin{figure}
      \resizebox{0.99\columnwidth}{!}{
	\includegraphics{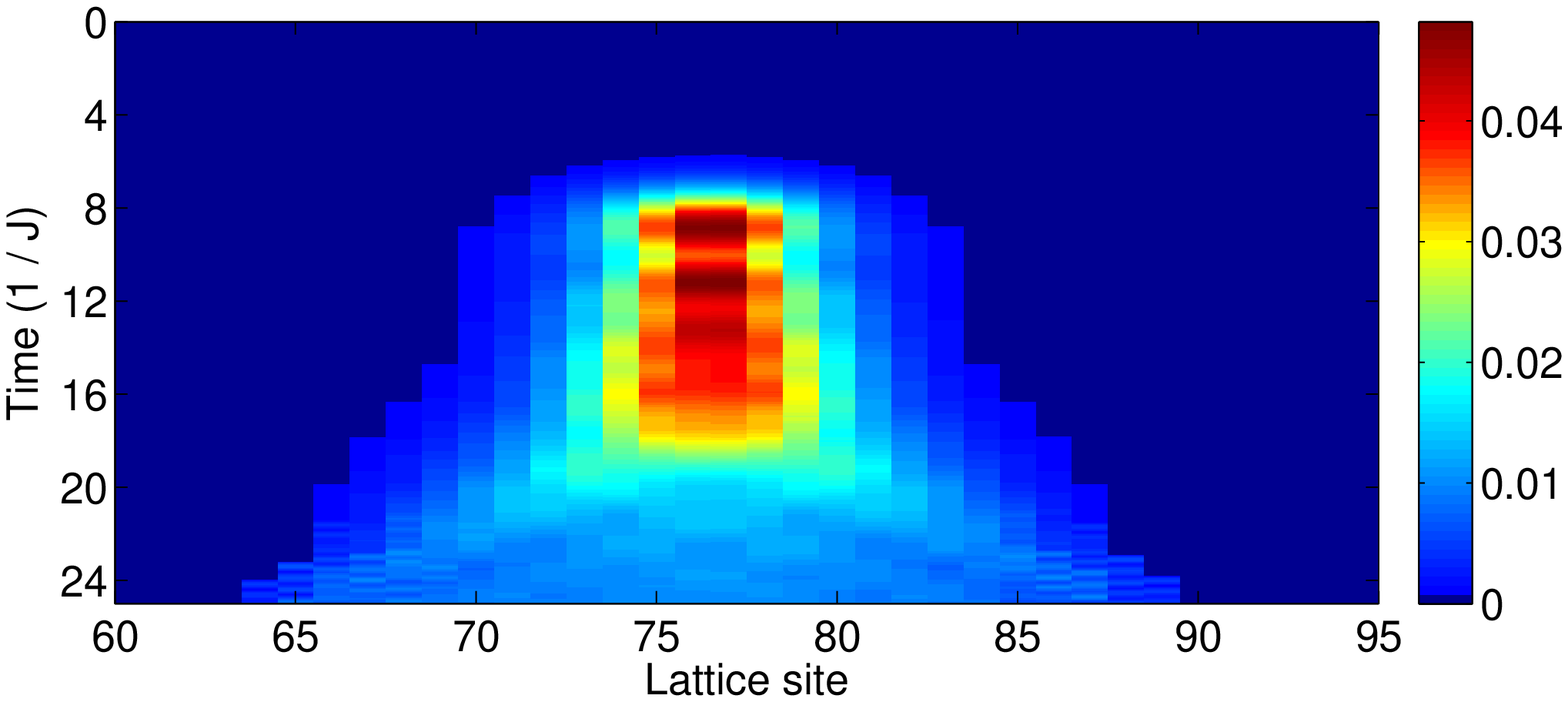}
      }
      \caption{The same as figure \ref{fig:pU1_imagesc}, but $|U| = 15.0$.}
      \label{fig:pU15_imagesc}   
    \end{figure}

    In Figures \ref{fig:pU1_imagesc} - \ref{fig:pU15_imagesc} we plot the square root of the density of the doublons
    $n_{\uparrow \downarrow, i} (t)$ for interactions $|U| = 1.0$, $|U| = 3.0$, and $|U| = 5.0$.
    Looking at Figures \ref{fig:pU1_imagesc} - \ref{fig:pU15_imagesc} we see that doublons are indeed initially created in the collision center, 
    and then they spread and possibly dissociate back to unpaired particles. Intriguingly, we see oscillations at the collision center.
    To examine the behaviour of the doublons better, let us plot the total number of doublons $n^{Total}_{\uparrow \downarrow} (t)$, given by

\begin{figure}
      \resizebox{0.99\columnwidth}{!}{
	\includegraphics{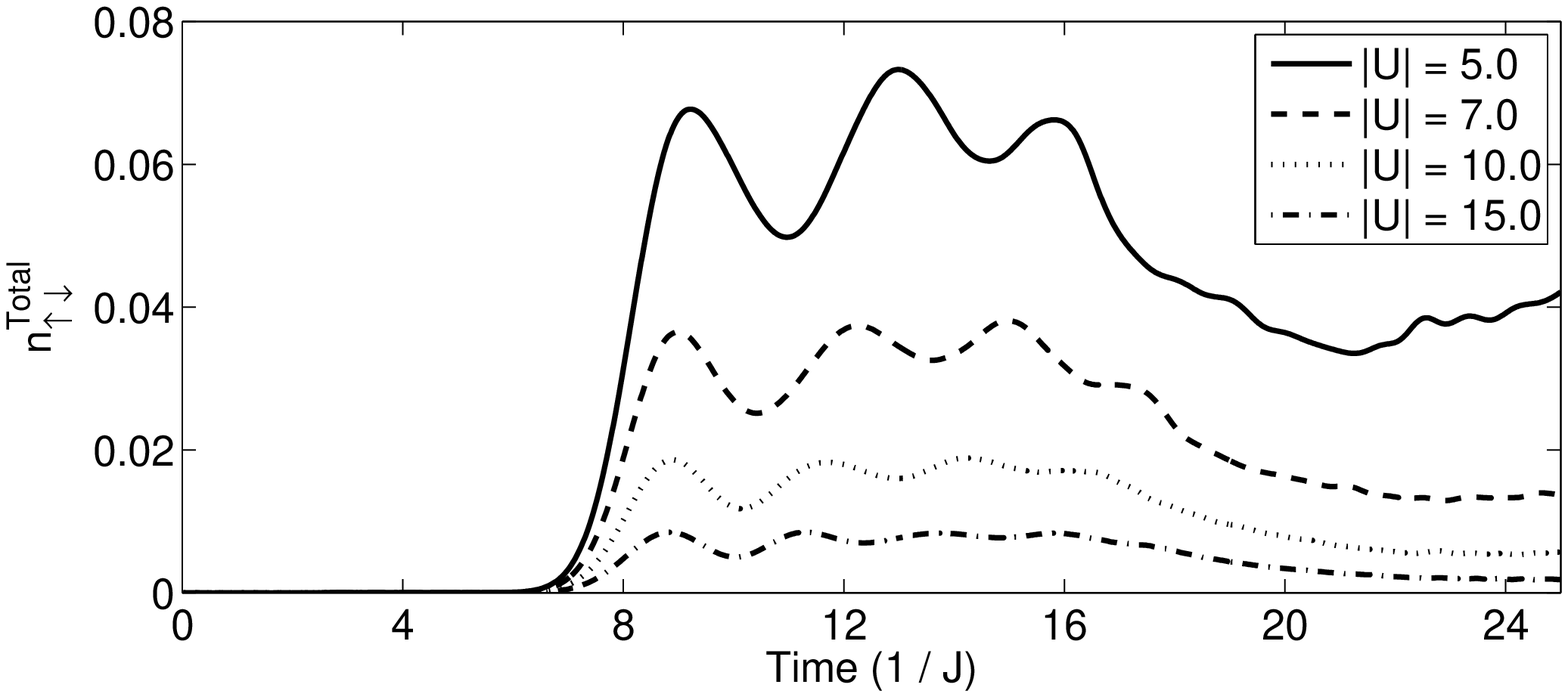}
      }
      \caption{$n^{Total}_{\uparrow \downarrow} (t)$ as obtained from TEBD numerics.}
      \label{fig:pairtotal}   
    \end{figure}

\begin{figure}
      \resizebox{0.99\columnwidth}{!}{
	\includegraphics{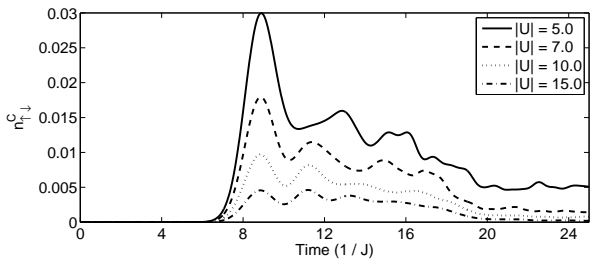}
      }
      \caption{$n^{C}_{\uparrow \downarrow} (t)$ as obtained from TEBD numerics.}
      \label{fig:pairC}   
    \end{figure}

    \begin{equation}
      \label{eq:totpairs}
      \begin{array}{ll}
	n^{Total}_{\uparrow \downarrow} (t) = \sum_{i=0}^L n_{\uparrow \downarrow, i} (t),
      \end{array}
    \end{equation}
    where $L$ is the lattice size $L = 150$. We also plot the total number of pairs in the collision center sites $n^{C}_{\uparrow \downarrow} (t)$, given by

    \begin{equation}
      \label{eq:Cpairs}
      \begin{array}{ll}
	n^{C}_{\uparrow \downarrow} (t) = n_{\uparrow \downarrow, L} (t) + n_{\uparrow \downarrow, R} (t),
      \end{array}
    \end{equation}
    where the sites L and R are the two dimer sites what we call ''the collision center'', see Figure \ref{fig:schematic}. 
    The quantities $n^{Total}_{\uparrow \downarrow} (t)$ and 
    $n^{C}_{\uparrow \downarrow} (t)$ are shown in Figures \ref{fig:pairtotal} and \ref{fig:pairC} for high interactions.

\begin{figure}
      \resizebox{0.99\columnwidth}{!}{
	\includegraphics{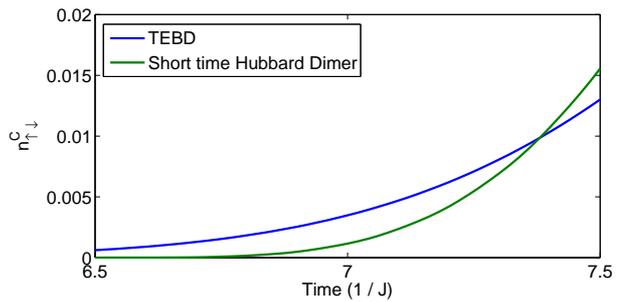}
      }
      \caption{Initial growth of $n^{C}_{\uparrow \downarrow} (t)$ as predicted by the Hubbard Dimer model compared to the TEBD numerics, $|U| = 5.0$.}
      \label{fig:comparisonpair5}   
    \end{figure}

\begin{figure}
      \resizebox{0.99\columnwidth}{!}{
	\includegraphics{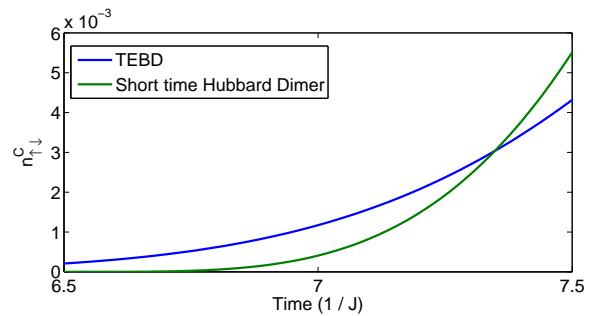}
      }
      \caption{Initial growth of $n^{C}_{\uparrow \downarrow} (t)$ as predicted by the Hubbard Dimer model, compared to the TEBD numerics, $|U| = 10.0$.}
      \label{fig:comparisonpair10}   
    \end{figure}

   Figures \ref{fig:pairtotal} and \ref{fig:pairC} tell about the dynamics of pair creation both at the collision center
   and in total. For short times just after the collision, the results seen in Figure \ref{fig:pairC} should be compared to Hubbard Dimer predictions, 
   c.f. Equation \eqref{eq:incoherent}. 
   These are shown in Figures \ref{fig:comparisonpair5} and \ref{fig:comparisonpair10} for $\frac{|U|}{J} = 5.0$ and $\frac{|U|}{J} = 10.0$.

    The plots in Figures \ref{fig:pairtotal} and \ref{fig:pairC} have been obtained from Equation \eqref{eq:incoherent}, assuming the density
    time-dependence to be the shape of a square pulse, i.e. $n^N_{un}(t) = 0.2$. In the numerics, the incoming cloud shape is not square but more like a Gaussian with 
    the height $0.3$ and the half-width at half maximum of $10$ lattice sites.  
    However, assuming a square shape simplifies the analysis significantly and is a reasonable approximation, since the 
    high density part of the cloud is anyway the one that contributes the most to the pair creation. 

    Elaborating on Figures \ref{fig:comparisonpair5} and \ref{fig:comparisonpair10}, in these Figures is shown the predicted growth of $n^{C}_{\uparrow \downarrow} (t)$
    from the beginning of the collision. 
    The differences between the predictions and the TEBD date are likely to be due to the assumption of
    a step function for the incoming unpaired density pulse. As said above, the actual incoming polarized cloud shape is more like a Gaussian,
    and thus we have neglected the small density increase in the beginning of the collision.
    Indeed, initially the step function underestimates the Gaussian, and at later times (when the short time approximation starts to break down)
    it overestimates it.

    Looking at Figure \ref{fig:pairtotal} one sees the beginning of the pair formation when the two clouds collide, at $t \approx 7 \frac{1}{J}$.
    The maximum in the pair density is reached after $t \approx 3 \frac{1}{J}$ from the start of the collision.
    Moreover, we see large scale oscillations in the pair density with a period $\approx 3 \frac{1}{J}$. 
    Now, since the Hubbard Dimer - based approximation \eqref{eq:incoherent} we considered above is valid
    only for times $t << \frac{\pi}{2} \frac{1}{J}$ we cannot explain the large scale oscillations or the maximum of amplitude using that model.
    Instead, we need to expand the short time considerations into longer times.  This is, as mentioned above, somewhat meticulous, but we hope
    to gain insight by formulating the general time-dependent equations although solving them might be difficult.

    In the case of longer times, we must take into account the fact that the dimer dynamics occurs in several other sites, not just at the collision center. 
    The dimer dynamics will occur at every site which has population of both up and down particles. When the collision progresses in time,
    more and more sites further away from the collision site will have have both up and down particles. Let us now define reaction center $R$ as all the lattice
    sites which have nonzero population of both up and down particles. Moreover, we define the reaction edge sites
    $Edge(t)$ to mean the last sites that have both up and down particles, when counting from the two sites of the collision center. 
    These edge sites change as a function of time as the collision progresses. 

    Now, unpaired particles will tunnel into the reaction center from the sites which are adjacent to the edge sites. The density of unpaired particles in these
    sites is denoted $n^N_{un}(t)$ analogously to the short time analysis. However, when longer times are considered, we need to take into account also unpaired
    particles tunnelling out of the reaction center. They tunnel out from the edge sites, at which the density is $n^{Edge}_{un}(t)$. Finally, in determining the
    total density of unpaired particles in the reaction center, we must consider that the unpaired particles will convert into doublons via Dimer dynamics.
    Summing up these contributions, we obtain for the total density of unpaired particles in the reaction center

    \begin{equation}
      \label{eq:un}
      \begin{array}{ll}

	\tilde{n}^{R}_{un} (t) = 2 * (sin(Jt))^2 n^{N}_{un} (t)\\
	- 2 * (sin(Jt))^2 n^{Edge}_{un} (t) - \tilde{n}^{R}_{\uparrow \downarrow} (t),

      \end{array}
    \end{equation}
    where $\tilde{n}^{R}_{\uparrow \downarrow} (t)$ is the total number of doublons in the reaction center. 
    Restating, the first term accounts for unpaired particles entering the reaction center, the second term accounts for unpaired particles leaving
    the reaction center and the last term accounts for unpaired particles converted into pairs.  Next we consider the doublons. With the definitions above,
    we hypothesise that the growth ($G$) and decay ($D$) of $\tilde{n}^{R}_{\uparrow \downarrow} (t)$ are given by:
    
   \begin{equation}
      \label{eq:growth}
      \begin{array}{ll}
	G (t) =  \int_{\tau = 0}^{\tau = t}  \int_{t^{'} = \tau}^{t^{'} = t} \frac{8}{16 + \frac{U^2}{J^2}} \\
	* (1 - cos(\sqrt{U^2 + 16J^2}(\tau-t^{'})) *  \tilde{n}^{R}_{un}(t^{'}) d\tau dt^{'},
      \end{array}
    \end{equation}

    \begin{equation}
      \label{eq:decay}
      \begin{array}{ll}
	D (t) =  \int_{\tau = 0}^{\tau = t}  \int_{t^{'} = \tau}^{t^{'} = t} \frac{8}{16 + \frac{U^2}{J^2}} \\
	* (1 - cos(\sqrt{U^2 + 16J^2}(\tau-t^{'})) *  \tilde{n}^{R}_{\uparrow \downarrow}(t^{'})d\tau dt^{'},
      \end{array}
    \end{equation}
    
    and the total number of doublons in the reaction center is

    \begin{equation}
    \label{eq:G-D}
      \begin{array}{ll}

	\tilde{n}^{R}_{\uparrow \downarrow} (t) = G(t) - D(t).

      \end{array}
    \end{equation}
    It is noted that as we have defined the reaction center consisting of all the sites which have nonzero population of up and down particles,
    the quantity $\tilde{n}^{R}_{\uparrow \downarrow} (t)$ is equal to $\tilde{n}^{Total}_{\uparrow \downarrow} (t)$ (unlike
    $\tilde{n}^{R}_{un} (t)$ which is not equal to $\tilde{n}^{Total}_{un} (t)$ since there exist unpaired particles outside the reaction center).
    Therefore, $\tilde{n}^{R}_{\uparrow \downarrow} (t)$ can be directly compared to 
    the total doublons densities seen in Figure \ref{fig:pairtotal} as a function of time (not
    to the two-site collision center doublon densities seen in Figure \ref{fig:pairC} that were considered above).
    The short time approximation we did initially involved 1) approximating the $\tilde{n}^{R}_{\uparrow \downarrow}$ term being neglibile
    in Equations \eqref{eq:un} and \eqref{eq:decay} 2) neglecting the tunnelling away of unpaired particles, i.e. the second term in Equation
    \eqref{eq:un} 3) considering that the relevant dynamics occurs at the two central cites, i.e. $Edge(t)$ = $L, R$.

\begin{figure}
      \resizebox{0.99\columnwidth}{!}{
	\includegraphics{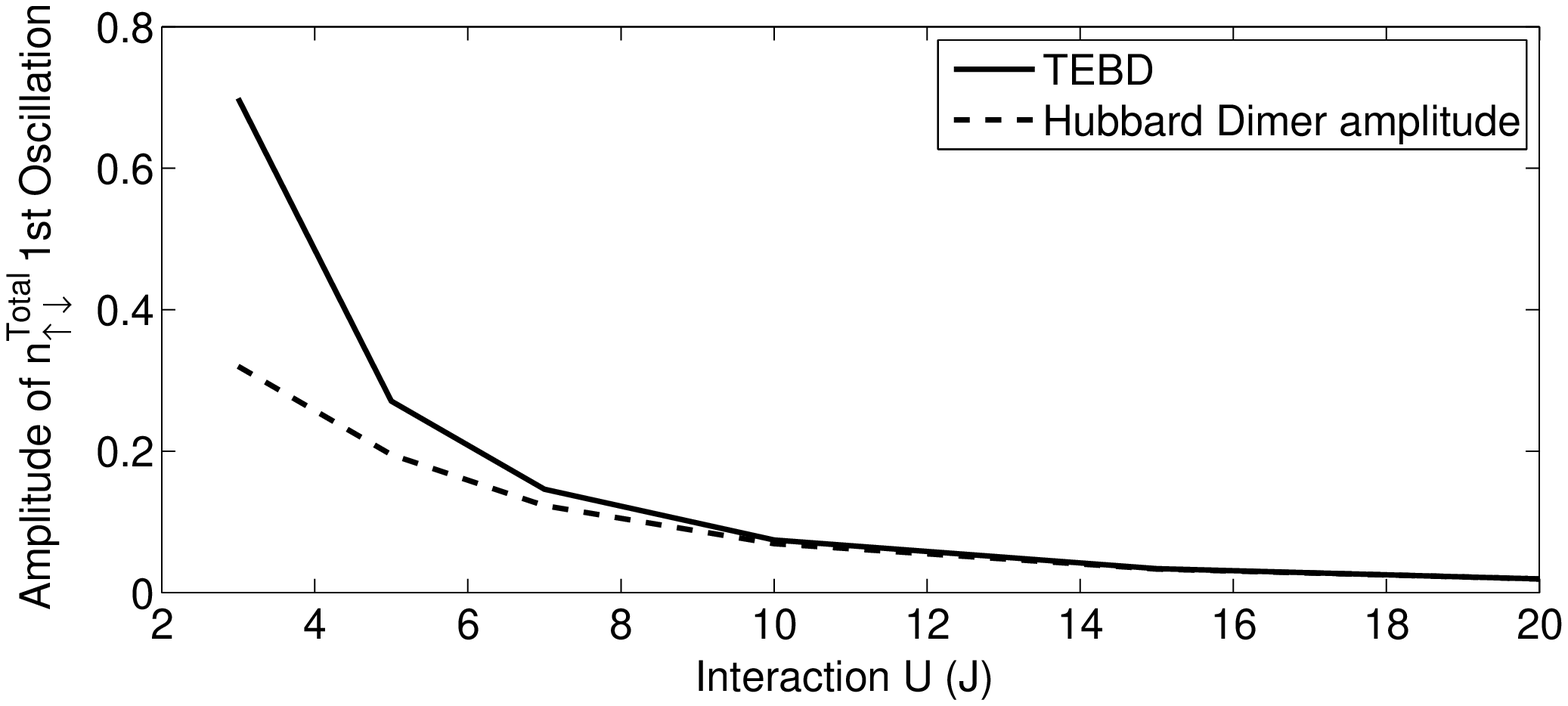}
      }      \caption{The height of the first peak of $n^{Total}_{\uparrow \downarrow} (t)$ seen in TEBD numerics fitted to $a * \frac{8}{16 + \frac{U^2}{J^2}}$.}
      \label{fig:amplitudecomp}   
    \end{figure}

    Equations \ref{eq:growth} - \ref{eq:G-D} could be solved self-consistently
    to obtain the full time evolution predicted by the model, but this is beyond the scope of this article. 
    Let us instead see if one learns something from the equations without
    solving them. In the high-interaction limit, we note that the cosine oscillations occur at such a high frequency that they average out.
    Thus, substituting $G(t)$ and $D(t)$ into Equation \ref{eq:G-D} and evaluating the $\tau$ integral:

    \begin{equation}
      \label{eq:growthfinal}
      \begin{array}{ll}
	\tilde{n}^{R}_{\uparrow \downarrow} (t) =  \int_{t^{'} = \tau}^{t^{'} = t} t * \frac{8}{16 + \frac{U^2}{J^2}} *  \tilde{n}^{R}_{un}(t^{'})dt^{'}.
      \end{array}
    \end{equation}  
    In the high interaction limit $|U| > 3$, $\tilde{n}^{R}_{\uparrow \downarrow} (t)$ in Equation \ref{eq:un} is neglibile, 
    as the Hubbard Dimer prefactor $\frac{8}{16 + \frac{U^2}{J^2}}$ 
    makes the density or doublons produced much less than the number of unpaired particles entering the reaction center from the expanding polarized clouds.
    Thus, in the high interaction limit we obtain:

    \begin{equation}
      \label{eq:finalresult}
      \begin{array}{ll}
	\tilde{n}^{R}_{\uparrow \downarrow} (t) =  \int_{t^{'} = \tau}^{t^{'} = t} t * \frac{8}{16 + \frac{U^2}{J^2}} \\
	* (2 * (sin(J t^{'}))^2 n^{N}_{un} (t^{'}) - 2 * (sin(J t^{'}))^2 n^{Edge}_{un} (t^{'}))dt^{'}.
      \end{array}
    \end{equation} 
    Which is our final result. 
    The density waves of expanding gases have the shape of a Gaussian, i.e. $n^{N}_{un} (t)$ and $n^{Edge}_{un} (t)$ are Gaussians. 
    To be more exact, the Gaussians have the mean $8 \frac{1}{J}$ (we start to count the time here at the beginning of the collision) and half width at half maximum $10$
    lattice sites (these, again, depend on the shapes of the polarized cloud). Therefore,
    based on the form of the equation \ref{eq:finalresult} we would expect to see in $\tilde{n}^{Total}_{\uparrow \downarrow} (t)$ 1) proportionality to 
    $\frac{8}{16 + \frac{U^2}{J^2}}$ as a function of interaction 2) Linear increase of amplitude superimposed on a wide Gaussian - shaped increase until 
    $8 \frac{1}{J}$ from the beginning of the collision after which
    wide Gaussian - shaped decrease. 3) Oscillations with period $\pi$, because the only oscillating term in the equation has the form $(sin(Jt))^2 = \frac{1 - cos(2Jt)}{2}$.
    This is promising since in the TEBD data in Figure \ref{fig:pairtotal} one sees oscillations at a period $\approx 3 \frac{1}{J}$ superimposed 
    with a Gaussian shape increase until $t = 6 \frac{1}{J} + 8 \frac{1}{J} = 14 \frac{1}{J}$. To examine whether the prediction (Equation \eqref{eq:finalresult})
    for the interaction dependence of the amplitude of these oscillations matches the TEBD numerics quantitaively, we plot
    the densities at the first visible oscillation peaks in Figure \ref{fig:pairtotal} 
    at $t=9 \frac{1}{J} - 10 \frac{1}{J}$ and fit the result to  $\frac{8}{16 + \frac{U^2}{J^2}}$.
    The result of the fit is shown in Figure \ref{fig:amplitudecomp}

    Intriguingly, we see that the fit in \ref{fig:amplitudecomp} is very good. 
    The fitting parameter $a$ was determined to be $a \approx 4.0$ which means that, in the long time limit,
    Hubbard Dimer dynamics indeed occur in several lattice sites close to the collision center.

\section{Conclusions}
\label{conclusions}
    
    We have simulated the collision of two polarized gases in 1D and in a lattice using TEBD numerics. 
    We found that there is $-U \leftrightarrow U$ symmetry in the collision
    and the gases bounce back from each other for interactions $|U| > 0.5$. Indeed, in 1D particles cannot pass each other without interacting. We propose that our
    analysis based on the Hubbard Dimer \cite{Kajala:2011p500}, which is a two site model, explains dynamically how pairs are created and dissociated during the collision. 
    This simple model explains why the gases bounce back: the Hubbard Dimer dynamics constrains the number of pairs that can be created in the time that the unpaired up and
    down particles are in contact during the collision. 
    Indeed, we compared the short time Hubbard Dimer analytical results to the numerics and found a good correspondence. In addition, we formulated time-development 
    for long times (self-consistent equations) and were able to identify the prominent features of long time pairing dynamics by examining the form of the equations.

    The derivation of the Hubbard Dimer dynamics does not importantyly include dimensionality dependence. It is possible that the same dynamics works for higher
    dimensions, as is suggested by our earlier work \cite{Kajala:2011p500} matching partly the results of a 2D experiment \cite{Schneider:2010p1468}.
    Interestingly, the simulations and analysis presented in this article could be mapped to the continuum case in the low density limit, 
    possibly relevant for the experiment \cite{Sommer:2011p2205}.
    in the \textit{quantum unitary evolution} regime;
    in \cite{Sommer:2011p2205} the dynamics of the collision was explained using semi-classical Boltzmann equations, which do not describe
    the low temperature regime. However, the Hubbard Dimer dynamics in the continuum limit is a subject of a further study. 
    Nonetheless, one can note the bouncing back - behaviour of the clouds, due to interaction, both in the experiment \cite{Sommer:2011p2205}
    and our simulations. Moreover, our results could give a quantitative prediction for the spin diffusivity at $T=0$. It should also be feasible
    to prepare Fermi gases in 1D lattice and in such sysytems our predictions could be directly tested.

    \begin{acknowledgement}

      We thank Andrew J. Daley and Jami J. Kinnunen for very useful discussions and feedback. 
      We acknowledge Mikko J. Leskinen for his help with numerics.
      This work was supported by the Academy of Finland (Projects No. 213362, No. 217043, No. 217045, No. 210953, and No. 135000) 
      and EuroQUAM/FerMix, and conducted (see www.esf.org/euryi) as a part of a EURYI scheme grant. 
      The research was partly supported by the National Science Foundation under Grant No. PHY05-51164. 
      Computing resources were provided by CSC - Finnish IT Centre for Science.

    \end{acknowledgement}
\bibliographystyle{unsrt}
%
%

\end{document}